\def\p{\partial}
\def\ext{\mathop{\rm Extr}}
\def\tr{\mathop{\rm Tr}}

\documentstyle[preprint,prl,floats,aps]{revtex}
\begin{document}
\draft

\title{Weight Space Structure and Internal Representations: a Direct Approach
to Learning and Generalization in Multilayer Neural Networks}

\author{R\'emi Monasson \cite{rm} and Riccardo Zecchina \cite{rz}}
\address{\cite{rm} INFN and Dipartimento di Fisica, P.le Aldo Moro 2,
I-00185 Roma, Italy\\
\cite{rz} INFN and Dip. di Fisica, Politecnico di Torino,
C.so Duca degli Abruzzi 24, I-10129 Torino, Italy}

\maketitle
\begin{abstract}

We analytically derive the geometrical structure of the weight space in
multilayer neural networks (MLN), in terms of the volumes of couplings
associated to the internal representations of the training set. Focusing on the
parity and committee machines, we deduce their learning and generalization
capabilities both reinterpreting some known properties and finding new exact
results.  The relationship between our approach and information theory as well
as the Mitchison--Durbin calculation is established.  Our results are exact in
the limit of a large number of hidden units, showing that MLN are a class of
exactly solvable models with a simple interpretation of replica symmetry
breaking.

\end{abstract}


\narrowtext

Memorization, rule inference or information processing by a neural
network may be seen as a complicated selection of one part of its
whole weight space \cite{cover,gard}.  Statistical mechanics has
permitted a quantitative study of this selection process for the
simple perceptron by providing a measure on the weight space resulting
from learning \cite{gard}. In particular, the purely geometrical
meaning of the spin-glass order parameter \cite{mpv} has been shown to
emerge naturally in this context.  These techniques have been
successfully applied to simple models of multilayer neural networks
(MLN) to compute their storage capacities and generalization errors
\cite{bhs,hertz,opper}.  However, a geometrical picture of MLN's weight
space and thus a unique ``conceptual'' frame allowing for the
interpretation of the physical and computational behaviour is lacking
so far.

In this letter we analytically derive such geometrical structure for MLN and
show how it is hidden in the usual Gardner's approach. The study of the
distribution of volumes of couplings associated to the internal
representations of the training set, leads to a simple geometrical
interpretation of replica symmetry breaking (RSB) and allows to deduce
the networks learning and generalization properties.  Moreover, we
show the key importance of the issue for analyzing the encoding of
information provided by the {\em internal representations}
\cite{bhs} in the intermediate layers of MLN by establishing a
correspondence with information theory and the Mitchison--Durbin
calculation \cite{mit}.  For the storage problem, we focus upon the
volumes giving the dominant contribution to Gardner's total volume,
whose number ${\cal N}_D$ is smaller than the total number ${\cal
N}_R$ of non--empty volumes. For the parity and committee machines
with $K(\gg 1)$ hidden units, ${\cal N}_D$ and ${\cal N}_R$ both
vanish at $\frac{\log K}{\log 2}$ and $\frac{16}{\pi}\sqrt{\log K}$
(so far unknown) respectively. Our results are shown to be exact in
this limit and are likely to coincide with the storage capacities of
both machines.  For finite $K$, we give a general geometrical
interpretation of RSB together with numerical results in the case
$K=3$.  The inference of a learnable rule is studied along the same
lines. We first reinterpret recent results\cite{opper} concerning the
Bayesian learning of a rule by a parity machine. We then explain the
smoothness of the generalization curve of the committee machine near
its Vapnik--Chervonenkis (VC) dimension \cite{vc} $d_{vc}\sim
\sqrt{\log K}$ and conjecture a cross-over to lower generalization
error for $\alpha \sim K$.

In the following, we shall consider tree-like MLN, composed of $K$
non-overlapping perceptrons with real-valued weights $J_{\ell i}$ and
connected to $K$ sets of independent inputs $\xi _{\ell i}$
($\ell =1,...,K$, $i=1,...,N/K$) \cite{bhs}.
The output $\sigma$ of the network is a binary function
$f(\tau _1 ,..., \tau _K)$ of the cells
$\tau_\ell  = \hbox{sign} ( \sum _i J_{\ell i} \xi _{\ell i})$ in the first
hidden layer.
The set $\{ \tau _\ell \}$ will be called hereafter {\em internal
representation} of the input pattern $\{ \xi _{\ell i} \}$.  For the
parity and committee machines, the decoder functions $f$ are
respectively $ \prod _\ell \tau _\ell $ and $\hbox{sign} ( \sum _\ell
\tau _\ell  )$.
The training set to be stored in the network includes $P=\alpha N$
patterns $\{\xi _{\ell i} ^{\mu}\}$ and their corresponding outputs
$\sigma ^{\mu}$ ($\mu =1,...,P$). For simplicity, both patterns and
outputs are drawn according to the binary unbiased distribution law.
In order to store the patterns, one must find a suitable
set of internal representations ${\cal T } = \{ \tau _\ell  ^{\mu} \}$
with a corresponding non zero volume
\begin{equation}
V_{ \cal T } = \int \prod_{\ell ,i} dJ_{\ell i} \prod_{\mu}
\theta \left( \sigma ^{\mu} f(\{\tau_\ell ^\mu\})
\right) \prod _{\mu ,\ell} \theta \left( \tau _\ell  ^{\mu} \sum _i
J_{\ell i} \xi _{\ell i} ^{\mu} \right)
\label{volume}
\end{equation}
where $\theta(\dots)$ is the Heaviside function and the integral over
the weights fulfills $\int \prod_{\ell ,i} dJ_{\ell i} = 1$.
Gardner's total volume is simply $V_G =\sum_{\cal T} V_{\cal T}$ and
the critical capacity of the network is the value $\alpha _c$ of the
maximal size of the training set such that $\overline{\log V_G}$ is
finite, where the bar denotes the average over the patterns and their
corresponding outputs \cite{gard}.  Moreover, the partition of $V_G$
into connected components may be naturally obtained using the $V_{\cal
T}$'s as elementary ``bricks''.  Indeed, from definition
(\ref{volume}), the set of weights $\{ J_{\ell i} \}$ contributing to
a given $V_{\cal T}$ is convex (or empty).  For the parity machine,
two volumes corresponding to two adjacent set of internal
representations (i.e. differing for one single $\tau _\ell ^{\mu}$)
cannot coexist (they would give opposite outputs for the pattern
$\mu$) and one of them at least must be empty. Thus each connected
component of $V_G$ coincides with one and only one volume associated to
an internal representation. For the
committee machine, a connected component of $V_G$ may include
several volumes $V_{\cal T}$.  The labelling of the different subsets
of $V_G$ using the internal representations of the training set ${\cal
T}$ may therefore be redundant depending on the particular decoder
under study. It is nevertheless a convenient starting point from the
analytical point of view and, as shown below, it does capture the main
features of the geometry of the coupling space.

The formalism recently introduced for a toy-model of MLN \cite{rete} can be
used to compute the distribution of the ``sizes'' of the volumes associated
to the internal representations ${\cal T}$. Once the canonical free-energy
$g(r)= - {1 \over Nr} \overline{\log ( \sum_{\cal T} V_{\cal T} ^
{\ ^{\displaystyle r}} )}$
is known, one obtains the micro-canonical entropy ${\cal N}(k)$ (i.e. the
logarithm of the typical number) of volumes $V_{\cal T}$ whose sizes are equal
to $k={1\over N} \log V_{\cal T}$ using the Legendre relations $k_r = {\p (r
g(r)) \over \p r}$ and ${\cal N}(k_r) = - {\p g(r) \over \p (1/r)}$
\cite{rete}.
The average over the patterns is performed using the replica trick for $r$
integer expecting that the final results remains valid for any real value of
$r$. There are $r$ blocks ($\rho=1,\dots,r$) of $n$ replicas ($a=1,\dots,n$).
Thus the spin glass order parameters  are the matrices ${\cal Q}_\ell $
and ${\hat {\cal Q}}_\ell $ of  the typical overlaps
$q_\ell ^{a\rho,b \lambda }= \frac{K}{N} \sum_i J_{i\ell }^{a \rho}
J_{i\ell }^{b \lambda } $ between
two weight vectors incoming onto the same hidden unit $\ell $ ($\ell
=1,\dots,K$) and of their conjugate Lagrange multipliers ${\hat q_\ell }^{a
\rho,b \lambda }$.
Since all the hidden units are indistinguishable, we assume
that at the saddle point ${\cal Q}_\ell ={\cal Q}$ and  ${\hat {\cal Q}}_\ell =
{\hat {\cal Q}}$ independently of $\ell $. Within the replica symmetric
(RS) Ansatz \cite{mpv}, we find
\begin{eqnarray}
g(r)=\ext _{q, q_*} \Bigg\{ && {1-r \over 2r} \log (1-q_*) -{1 \over 2r}
\log(1-q_*+r(q_*-q)) - {q \over 2 (1-q_*+r(q_*-q))} \nonumber \\
&&-{\alpha \over r } \int \prod _\ell Dx_\ell\ \log {\cal H}(\{ x_\ell \})
\Bigg\}
\label{g_r}
\end{eqnarray}
where ${\cal H}(\{ x_\ell \})=\tr _{\{\tau _\ell \}} \prod _\ell
\int Dy_\ell H[(y_\ell \sqrt{ q_* - q } + \tau _\ell  x_\ell
\sqrt{q})/ \sqrt{ 1-q_* }]^ r$.
Here, $q_*(r)=q^{a\rho,a \lambda}$ and $q(r)=q^{a \rho,b \lambda}$ are the
typical overlaps between two weight vectors corresponding to the same
($a$,$\rho \ne \lambda $) and
to different ($a\ne b$) internal representations ${\cal T}$ respectively
\cite{gard,rete}. The Gaussian measure is denoted by $Dx=\frac{1}{\sqrt{2 \pi}}
e^{-x^2/2}$ whereas the function $H$ is defined as $H(y)=\int_y^\infty Dx$.
In eqn.(\ref{g_r}), the sum $\tr _{\{\tau _\ell\}}$ runs
over the internal representations $\{ \tau_\ell\}$
giving a positive output $f(\{\tau_\ell\})=+1$ only, since the
outputs $\sigma ^{\mu}$ can always be set equal to $+1$ at the cost of
redefining the input patterns.

The whole distribution of sizes is available through $g(r)$.
When $N \to \infty$, $\frac{1}{N} \overline{
\log(V_G)}=-g(r=1)$ is dominated by volumes of size $k_{r=1}$ whose
corresponding entropy (i.e the logarithm of their number divided by $N$) is
${\cal N}_D = {\cal N}(k_{r=1})$. At the same time the most numerous volumes
are those of smaller size
$k_{r=0}$, since in the limit $r \to 0$ all the ${\cal T}$ are
counted irrespectively of their relative volumes.
Their corresponding entropy ${\cal N}_R = {\cal N}(k_{r=0})$ is
the normalized logarithm of the total number of implementable internal
representations.
The quantities ${\cal N}_D$ and ${\cal N}_R$ (that for lack of space we do not
write explicitly) are easily obtained from the RS free--energy eqn.(\ref{g_r})
using the above Legendre identities. In particular, $q(r=1)$ is the usual
saddle point overlap of the Gardner volume $g(1)$ \cite{gard,bhs}.
The vanishing condition for the entropies should coincide with the zero
volume condition for $V_G$ and thus should give the storage capacity of the
models.

Both ${\cal N}_D$ and ${\cal N}_R$ have a straightforward interpretation in the
context of information theory. One can easily verify that the quantity of
information ${\cal I}$  carried by the distribution of the implementable
internal representations ${\cal T}$ about the weights, $\displaystyle{ {\cal
I}=-\sum_{{\cal T}} \frac{V_{\cal T}}{V_G} \log \frac{V_{\cal T}}{V_G} }$ , is
equal to ${\cal N}_D$. The information capacity, i.e. the maximal quantity of
information one can extract from the internal representations, is achieved when
all internal representations ${\cal T}$ are equiprobable and thus equals ${\cal
N}_R$. One should notice that the Mitchison--Durbin \cite{mit} geometrical
calculation is simply an upper (and decoder--independent) bound on ${\cal
N}_R$.

Let us see now the physical and geometrical interpretation of ${\cal N}_D$.
Fig.~1 displays the RS entropy ${\cal N}_D$ as a
function of $\alpha$ for both the parity and committee machines with
$K=3$ hidden units. This entropy vanishes at a critical value $\alpha
_D$ of the size of the training set.  Numerically, we find $\alpha _D
\simeq 3.8$ and 2.9 for the parity and the committee machines
respectively. For comparison,  the storage capacities
obtained with the one step RSB Ansatz are $\alpha _c \simeq 5$ and 3
respectively \cite{bhs}. Being the entropy of a discrete system,
${\cal N}_D$ cannot be negative and therefore $\alpha _D$ is an upper
bound of the size of the training set $\alpha_{RSB}$ where the replica
symmetry breaking occurs for both ${\cal N}_D$ and $V_G$ \cite{rete} .
It is indeed known that $\alpha _{RSB} = 3.2$ and 1.8 for the parity
and the committee machines respectively \cite{bhs}. When $\alpha <
\alpha _{RSB}$, the RS assumption is exact whereas ${\cal N}_D$ is positive,
showing that the number of internal representations volumes
contributing to $V_G$ is exponentially large with $N$.
$q_*$ measures the typical overlap inside one of
these volumes, while the usual overlap $q$ arising in the RS
computation of $V_G$ tells us how far away are two different volumes
$V_{\cal T}$.  The behaviour of $q_*$ versus $\alpha$ is shown in the
inset of figure 1.  When choosing randomly two weights vectors storing
the training set, the probability that they belong to the same
$V_{\cal T}$ vanishes as $\exp(-N {\cal N}_D)$ and their overlap
distribution cannot be told from a Dirac peak in $q$, as must be for
the RS solution to be exact.  As a consequence, the blind computation
of $V_G$, though it gives correct results, hides the
geometrical structure of the weight space. In the limit of a
large number $K$ of hidden units, the asymptotic expressions of the
overlaps and of $\alpha _D$ may be obtained analytically.  We find
that $q=0$ and $q \simeq 1-\frac{128}{\pi ^2 \alpha^2}$ for the parity
and the committee machines respectively and that $q_* \simeq 1
-\frac{\pi ^2 \Gamma ^2}{2 \alpha ^2 K^2}$ in both cases with $\Gamma
=-1/(\sqrt{\pi} \int du H(u) \log H(u)) \simeq 0.62$.  The
corresponding entropies ${\cal N}_D^{(Par)}\simeq \log K -\alpha \log
2$ and ${\cal N}_D^{(Com)}\simeq \log K -\frac{\pi ^2 \alpha ^2}{256}$
vanish at $\alpha _D ^{(Par)} \simeq \frac {\log K}{\log 2}$ and
$\alpha _D ^{(Com)} \simeq \frac{16}{\pi} \sqrt{
\log K}$.

When $\alpha > \alpha_{RSB}$, the computation of ${\cal N}_D$ requires the
introduction, at the first stage of RSB, of four order parameters
$q_*',q_0,q_1,m$~: $q_*'$ is the internal overlap of the internal
representations volumes and $q_0,q_1,m$ are simply the usual parameters arising
in the one step Gardner's computation \cite{futuro}.
For brevity we only present below our
numerical results together with their geometrical interpretation.
Above $\alpha _{RSB}$, there exist a finite number of big
regions with mutual overlap $q_0$. Each region $\rho$ contains an exponential
number of volumes $M_{\rho}$ of internal overlap $q_*'$ and typically
separated by an overlap $q_1$. The number of such regions may be
roughly estimated by $\frac{1}{1-m}$, since $m=1-\sum_{\rho}
(M_{\rho}/\sum _{\rho'} M_{\rho'} )^2$, whereas in the RS phase $m=1$.
We have checked numerically this geometrical scenario for the parity
machine with $K=3$ hidden units (numerically much simpler than the
committee machine case since $q_0=0$ at the saddle point).  The
internal overlap $q_*'$ is continuous at the RSB transition -- see the
inset of fig.~1 -- with $q_*<q_*'$ for $\alpha > \alpha_{RSB}$. We
conjecture that increasing $\alpha$ a whole continuous breaking of RSB
occurs. The geometrical process should then be thought of as a
progressive shrinking and disappearance of volumes
with internal overlap $q_*(\alpha)$ inside
sub-regions characterized by $q(x,\alpha)$ \cite{mpv}.  In fig.~1, we
have reported the curve of ${\cal N}_D$ computed with this one step
Ansatz for the parity machine $K=3$. $\alpha_D$ increases
from $\simeq 3.8$ (RS value) to a value close to $5$ and thus to the
one step RSB value of $\alpha_c$\cite{bhs} (since $q_*'$ and $q_1$
are close to 1, our numerical results become less precise for
$\alpha$ larger than $\sim 4.1$ and $\alpha_D \simeq 5$ is obtained
through the linear extrapolation corresponding to the dashed part of
the curve).

The RS calculation of ${\cal N}_R$ for both machines, leads the
following general results.  When $\alpha
< {2\over K}$, one finds that all the $2^{(K-1)P}$ internal
representations may be implemented. This obviously coincides with the storage
capacity of the hidden perceptrons seeing only $N/K$ input units. For
$\alpha > {2 \over K}$, we find that at the saddle-point $q_*=1$,
meaning that the most numerous volumes $V_{\cal T}$ are almost empty and are
therefore the smallest ones at the same time. The resolution of the
saddle-point equations requires the introduction of a new order parameter
$\mu =\displaystyle{ \mathop{\hbox{lim}}_{r \to 0}\  r / (1-q_*)}$,
describing how quickly the typical size of the volumes decreases with respect
to the inverse ``temperature'' $r$ \cite{rete}. For the parity machine
with $K \ge 3$, $q=0$, $\mu =\alpha K(\alpha K -2)$ is always a locally
stable saddle-point giving ${\cal N}_R^{(Par)} = \alpha K \log (\alpha K)
 - (\alpha K -1) \log ( \alpha K-1) - \alpha \log 2$
which exactly saturates the upper bound derived by
Mitchison--Durbin \cite{mit}. In the case of the committee machine,
a simple analytical expression for ${\cal
N}_R^{(Com)}$ is not available for finite $K$.  Once more in fig.~1,
we report the numerical results concerning the RS calculations of
${\cal N}_R$ for both machines with $K=3$.  The value $\alpha_R$
at which ${\cal N}_R$ vanishes should satisfy the obvious
inequality $\alpha_D \leq \alpha_R \leq \alpha_c$; the RS
approximation however overestimates $\alpha_R$ leading to an
expression which is slightly larger than the one step value of
$\alpha_c$.  For the parity and committee machines with $K=3$ we find
$\alpha_R=5.4$ and $3.5$ respectively. This is an evidence for the
necessity of RSB to compute exactly ${\cal N}_R$ for finite $K$.

When $K\gg 1$,  ${\cal N}_R$ (resp. $\alpha _R$) is asymptotically equal
to ${\cal N}_D$ (resp. $\alpha _D$).
In the case of the parity machine $\alpha_D$ and $\alpha_R$ also coincide
with the known value of $\alpha_c=\frac{\log K}{\log 2}$\cite{bhs}.
We expect the same equality ($\alpha_D =\alpha_R =\alpha_c=\frac{16}{\pi}
\sqrt{\log K}$) to hold in the case of the committee machine.
In order to show that the RS solution of ${\cal N}_R$ is asymptotically
correct, we have checked its local stability with respect to fluctuations of
the order parameter matrices.  Although it would require a complete analysis of
the eigenvalues of the Hessian matrix, we have focused only on the
replicons 011 and 122 in the notations of \cite{kondor}, which are usually
the most ``dangerous'' modes \cite{mpv}. For a free--energy functional
depending only on one order parameter matrix $q^{a\rho ,b\lambda}$,
the corresponding eigenvalues are $\Lambda _{011}$ and $\Lambda
_{122}$ given by formula (41) in ref.\cite{kondor}.  In our case,
however, the free-energy depends on $2K$ matrices $\{ {\cal
Q}_l,{\hat{\cal Q}}_l \}$ and the stability condition for each mode
reads
$\Delta (\alpha ,K) = \hat \Lambda \ (\  \Lambda + (K-1) \overline{
\Lambda}\ ) - {1 \over K^2} < 0$
where $\hat \Lambda,\Lambda,\overline{\Lambda}$ are the eigenvalues
computed for the fluctuations with respect to
${\hat{\cal Q}}_\ell  {\hat{\cal Q}}_\ell $, ${\cal Q}_\ell  {\cal Q}_\ell $
and
${\cal Q}_\ell {\cal Q}_m$ ($\ell \ne m$) respectively \cite{gard,bhs}.
A tedious calculation leads to the final expressions $\Delta_{011}$
and $\Delta_{122}$ \cite{futuro}. For the parity machine, we find
$\Delta _{011} ^{(Par)} (\alpha , K)= {\alpha \over K} ( {2\over \pi}
+ {1\over \alpha K} ( 1 - {4\over \pi})) ^2 -
{1 \over K^2}$ and $\Delta _{122} ^{(Par)} (\alpha , K) =0$
which are valid for $K\ge 3$ and $\alpha\ge \frac{2}{K}$. The RS solution
is unstable against 011 replicon mode for $\alpha\ge \frac{3.27}{K}$ (i.e. of
the same order as the storage capacity of each single input perceptron).
However, in the large $K$ limit, $\Delta_{011}$ vanishes. For the
committee machine, one finds
$\Delta _{011} ^{(Com)} (\alpha , K) \simeq {\sqrt{2} \over \pi ^3 K}$
and $\Delta _{122} ^{(Com)} (\alpha , K) \simeq - {1 \over 2 K^2}$
for $K\gg1$ and $\alpha\gg1$. We notice that the 122 mode is always stable and
a unique order parameter $q_*$ is thus sufficient to describe the volume
associated to a set of internal representations ${\cal T}$.
For both machines, our RS solution is marginally stable when $K\to\infty$
and should therefore become exact in this limit.

In order to understand what are the consequences of the weight space structure
on the generalization ability of MLN, we now modify our approach to the case of
deterministic input--output mappings.

The case of the parity machine trained on a learnable rule
(i.e. generated by a ``teacher'' network endowed with an identical
architecture) has been recently studied
\cite{opper} in the Bayesian framework where the
generalization properties are derived through the knowledge of the entropy
$S_G=-\frac{1}{N}\overline{ V_G \log V_G}$.
The transition from high generalization error $\epsilon _g=\frac{1}{2}$ to
low $\epsilon _g$(=$\frac{\Gamma}{\alpha}$ for large $\alpha$)
\cite{opper} may be geometrically understood along the lines
developed above. The free-energy $s(r)$ generating the distribution of the
``sizes'' of the internal representation volumes $V_{\cal T}$ becomes
$s(r)= - {1 \over Nr}\ \overline{ \sum_{\cal T} V_{\cal T} ^
{\ ^{\displaystyle r}}\log(\sum_{\cal T}V_{\cal T} ^{\ ^{\displaystyle r}})}$
where we obviously recover $S_G=s(1)$. The replica calculation
of $s(r)$ technically differs from the computation of $g(r)$ by taking
the limit $n\to 1$ instead of $n\to 0$ \cite{futuro}. Within the RS Ansatz,
we find
\begin{eqnarray}
s(r)=\ext_{q} \Bigg\{ &&{1-r\over 2r}\log(1-q_*)-{1\over 2r}\log(1-q_*+r(q_*
-q))-{q\over 2(1+(r-1)q_*)} \nonumber \\
&&- {2\alpha\over f(q_*)}\int \prod _\ell Dx_\ell {\cal H}(\{ x_\ell \})
\log {\cal H}(\{ x_\ell \}) \Bigg\}
\label{s_r}
\end{eqnarray}
with $f(q_*)=[2\int Dz H(z\sqrt{q_*}/\sqrt{1-q_*})^r]^K$ and $q_*(r)$ is the
saddle-point overlap of $s_0(r)=(1-r)\log(1-q_*)-\log(1+(r-1)q_*)-2\alpha
\log f(q_*)$.
The logarithm ${\cal M}_D$ of the number of the internal
representation volumes contributing to the Bayesian entropy $S_G$ is
given by ${\cal M}_D=\frac{\partial s}{\partial r}(r=1)$.
We find that for $\alpha K \gg 1$, $q_*\simeq 1-\frac{\pi ^2 \Gamma ^2}
{2\alpha ^2 K^2}$.
In the case of the parity machine $q=0$, ${\cal M}_D ^{(Par)}
\simeq \log K -\alpha \log 2$ for $\alpha <\alpha _0= \frac{\log K}{\log 2}$
and $q=q_*$, ${\cal M}_D=0$ for $\alpha > \alpha _0$. Thus, below
$\alpha _0$, the weight space is composed of an exponentially large
number of volumes and the typical overlap $q$ between the volume
occupied by the teacher and any other one is zero~: $\epsilon_g={1\over 2}$.
Above $\alpha _0$, since only one internal representation survives,
the student has fallen down into the teacher volume~: $q=q_*$ and
$\epsilon_g\simeq \frac{\Gamma}{\alpha}$. When $\alpha <\alpha _0$,
$S_G ^{(Par)}=\alpha \log 2$, meaning that all the sets of $P$
outputs are equiprobable.  Choosing them with a probability
$V_G(\{\sigma\})$ is then equivalent to drawing them randomly.
This is the reason why $\alpha _D$
defined for the storage problem (and more generally $d_{vc}$) appears
on the generalization curve of the parity machine.  Our calculation
also indicates that the computation of $\alpha_0$ should include RSB
effects for finite $K$, while the asymptotic RS expression of
$\epsilon_g$ ought to be exact, as has been found for the
non--monotonic perceptron \cite{rete2}.

Turning to the committee machine, a calculation of the Bayesian entropy
$S_G$ similar to \cite{hertz} leads to the following results when
$K\gg\alpha\gg 1$. The typical teacher--student overlap $q$ decreases as
$1-\frac{\pi ^6\Gamma ^4}{2\alpha ^4}$ giving
an entropy $S_G^{(Com)} \simeq 2\log \alpha$ and $\epsilon_g\simeq\frac{
2\Gamma}{\alpha}$. This shows that, at variance with the parity machine case,
only a small fraction among the $2^P$ possible sets of outputs contribute
to $S_G^{(Com)}$ and explains why the generalization curve is smooth
for $\alpha \simeq \sqrt{\log K}$ (which is the order of magnitude of
$d_{vc}$). We find ${\cal M}_D ^{(Com)} \simeq \log K - \log
\alpha $, confirming that $\alpha _D$ (and thus $d_{vc}$) is not relevant
to the computation of the typical generalization error.
At $\alpha _{c.o.}\sim K$, only a single internal representation subsists
and beyond this critical size of the training set the generalization
error should equal $\epsilon_g=\frac{\Gamma}{\alpha}$ as is for finite $K$ and
large $\alpha$ \cite{hertz}. Note that the order of magnitude of $\alpha
_{c.o.}$ is corroborated by the condition $q=q_*$ one has to fulfill once a
unique $V_{\cal T}$ remains non--empty. A rigorous proof of the presence of
this cross--over (from $\epsilon_g=2 \frac{\Gamma}{\alpha}$ to
$\epsilon_g=\frac{\Gamma}{\alpha}$) at $\alpha_{c.o.}$
would however require to extend the validity of our
calculation to the regime $1\ll \alpha \sim K$.

We are grateful to N. Brunel, M. Budinich, M. Ferrero and D. O'Kane for
discussions.

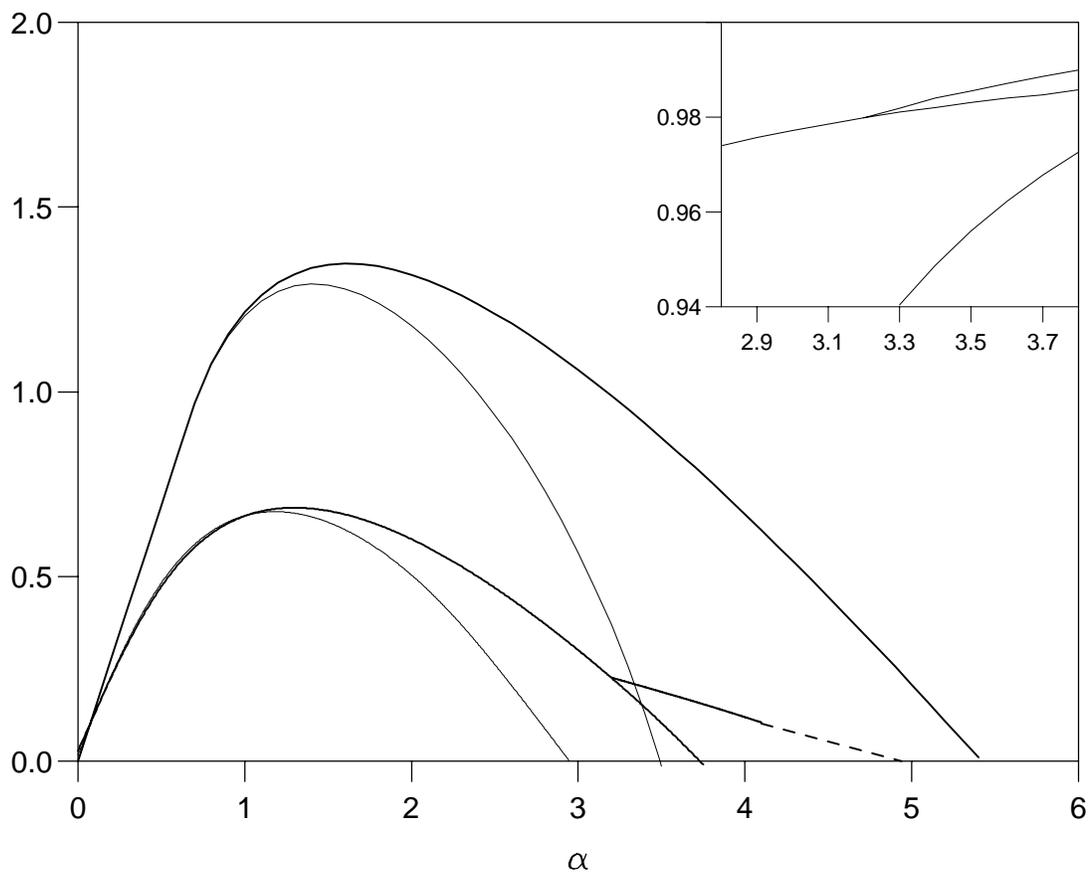
\begin{figure}
\caption{${\cal N}_R$ (upper curves) and ${\cal N}_D$ (lower curves) for the
parity machine (bold) and the committee machine (light), with $K=3$ hidden
units. Inset: $q_1,q_*,q_*'$ (lower, middle and upper curves
respectively) versus $\alpha$ for the parity machine ($q_1$ starts at
$\alpha=\alpha_{RSB} \simeq 3.2$ with a value close to $0.93$).}

\label{fig1}
\end{figure}

\end{document}